\documentclass[epj]{svjour}
\usepackage{epsfig}
\usepackage{amsfonts}
\usepackage{amsmath}
\newcommand{\s}[1]{{\rlap/ #1}}

\begin{document}
\title{
Forward-like functions for dual parametrization of GPDs from the nonlocal
chiral quark model}
\author{Kirill M. Semenov-Tian-Shansky \inst{1} \inst{2}
\thanks{\emph{semenovk@tp2.ruhr-uni-bochum.de} }%
}                     
%
%
\institute{Institut f\"{u}r Theoretische Physik II, Ruhr-Universit\"{a}t Bochum, D-44780 Bochum, Germany
\and St. Petersburg State University, St.Petersburg, Petrodvoretz, 198504, Russia }
\date{Received: 17 March 2008 / Revised version: 28 April 2008}
%
\abstract{
We derive the set of inversion relations allowing
to establish the link between the dual parametrization
of GPDs and a broad class of phenomenological models
for GPDs. As an example we consider the results of
the calculation of the pion GPD in the nonlocal
chiral quark model (NlCQM) to recover the set of forward-like
functions
$Q_{2 \nu}$
representing this GPD in the framework of dual parametrization.
We also argue that Abel tomography method overlooks
possible $\delta$-function like contributions to
GPD quintessence function which make explicit contribution
to the $D$-from factor.
} 

\authorrunning{K. Semenov-Tian-Shansky }
\titlerunning{Forward-like functions for dual parametrization of GPDs from NlCQM }

\maketitle
%

\section{Introduction}

Generalized parton distributions (GPDs)
\cite{pioneers}
have been in focus of intensive studies during the past decade.
These distributions, which arise as natural generalization
of parton distribution functions (PDFs) familiar from inclusive reactions,
proved to be an extremely efficient tool for the description
of hadron properties in terms of quark and gluonic degrees
of freedom. Experimentally GPDs are accessed through
hard exclusive reactions. Reviews of both theoretical and
experimental aspects are given {\it e.g.} in refs
\cite{GPV,Diehl,BelRad,Boffi}.

It is needless to say that the direct extraction of GPDs
from the amplitudes of hard exclusive processes is highly
demanded, since these functions contain a wealth of information
on hadron structure. Unfortunately, the problem turns out to
be very complicated, since GPDs depend on three variables
($x,\,\xi, \, t$)
as well as on renormalization scale.
Moreover, GPDs always enter the observable quantities
(cross-sections, spin asymmetries {\it etc.})
as certain convolutions with perturbative kernels.
All this makes the extraction of GPDs from
the observables an extremely difficult task. As a palliative
different models for GPDs are employed.
These models should satisfy the requirements of
polynomiality and positivity which are the
consequences of general principles of Lorentz invariance and
positivity of probability.

Historically the most popular model for GPDs is double distribution
parametrization suggested by Radyushkin
\cite{RadDDandEvolution}.
In this parametrization the GPD is given as a convolution of
a forward parton distribution
$q(\beta)$
with a specific profile function
$h(\beta, \alpha)$:
\begin{equation*}
\begin{split}
&H^q(x,\xi, t=0)  = \\& \int_{-1}^1 d \beta
\int_{-1+| \beta|}^{1-|\beta|} d \alpha \,
\delta(x- \beta - \alpha \xi) \, h(\beta, \alpha) q(\beta).
\end{split}
\end{equation*}
This parametrization presents a simple and convenient
form of GPDs and is used in numerous phenomenological applications.
However, as  was pointed out in
\cite{Polyakov:1999gs},
it does not satisfy the polynomiality condition in its full form.
To overcome this problem one has to complete GPD in double parametrization
with the so-called $D$-term.

A different elegant way to implement polynomiality of GPDs
was proposed in
\cite{Polyakov:2002wz}.
This alternative parametrization is called dual since it is based on
the representation of GPDs as infinite series of $t$-channel exchanges.
In this paper we discuss some properties of dual parametrization
and derive the set of inversion relations allowing to establish the
link between a broad class of phenomenological models of GPDs and the dual
parametrization. As an example we consider the nonlocal chiral quark
model for quark GPDs in a pion and with
the help of the inversion formulas determine
the shape of  the corresponding forward-like functions
$Q_0$, $Q_2$, $Q_4$.
We compare these functions to the GPD quintessence function
$N(x)$
calculated with the help of the method of
Abel tomography
\cite{Tomography}.
We also calculate the $D$-form factor as well
as the first coefficient of Gegenbauer expansion of the
$D$-term in the framework of nonlocal chiral quark
model.

\section{Basic facts on the dual parametrization of GPDs}
\label{basic_facts}

Below we consider only the case of the singlet
($C=+1$)
GPD for spin-$0$ target. In the forward limit
this GPD is reduced to
$\frac{1}{2}(q(x)+\bar{q}(x))$.
The generalization for the nonsinglet
($C=-1$) GPD
(the one reducing to
$\frac{1}{2}(q(x)-\bar{q}(x))$
in the forward limit)
is straightforward.

According to the result of
\cite{Polyakov:1998ze},
the partial wave decomposition in the $t$-channel for singlet GPD
$H(x, \xi, t)$
can be written as the following formal series:
\begin{equation}
\begin{split}
& H(x, \xi, t)= \\&
\sum_{n=1 \atop \text{odd}}^\infty
\sum_{l=0 \atop \text{even}}^{n+1}
B_{nl}(t) \,
\theta
\left(
1-\frac{x^2}{\xi^2}
\right)
\left(
1-\frac{x^2}{\xi^2}
\right)
C_n^{\frac{3}{2}}
\left( \frac{x}{\xi} \right)
P_l
\left( \frac{1}{\xi} \right),
\label{FSGPD}
\end{split}
\end{equation}
where
$C_n^{\frac{3}{2}}(\chi)$
stand for Gegenbauer polynomials;
$P_l(\chi)$
are Legendre polynomials and
$B_{nl}(t)$
are the generalized form factors;
$x$, $\xi$
and
$t$
stand for usual GPD variables.

The problem of summing up the formal series
(\ref{FSGPD})
was solved in
\cite{Polyakov:2002wz}.
For this one has to introduce
the set of {\em forward-like functions}
$Q_{2 \nu}(x,t)$ $(\nu=0, \,1,\,...)$
whose Mellin moments generate the generalized form factors
$B_{nl}(t)$:
\begin{equation}
 B_{n \, n+1-2 \nu}(t)= \int_0^1 dx x^n Q_{2 \nu}(x,t)\,.
\label{BnL_def}
\end{equation}
The set of forward-like functions
$Q_{2 \nu}(x,t)$
is connected to GPD
$H(x,\xi,t)$
in dual parametrization with the help of the following relation:
\begin{equation}
\begin{split}
& H(x,\xi,t)=
\sum_{\nu=0}^\infty
\left\{
\frac{\xi^{2 \nu}}{2} 
\left[
H^{( \nu)}(x,\xi,t)-H^{( \nu)}(-x,\xi,t)
\right]\,
\right.
\\&
-\left(
1- \frac{x^2}{\xi^2}
\right)
\theta(\xi-|x|) \\&  \times
\sum_{l=1 \atop \text{odd}}^{2 \nu -3}
C_{2 \nu-l-2}^{\frac{3}{2}}
\left( \frac{x}{\xi} \right)
\left.
P_l \left( \frac{1}{\xi} \right)
\int_0^1 dy y^{2 \nu-l-2} Q_{2 \nu}(y,t)
\right\}
\, ,
\end{split}
\label{H_dual_through_Qk}
\end{equation}
where the functions
$H^{( \nu)}(x, \xi, t)$
defined for
$-\xi \le x \le 1$
are given by the following integral transformations:
\begin{equation}
\begin{split}
 H^{(\nu)}(x,\xi,t)&=
\theta(x>\xi)
\frac{1}{\pi}
\int_{y_0}^1 \frac{dy}{y}
\left[
\left(
1-y \frac{\partial}{\partial y} \right) Q_{2 \nu}(y,t)
\right] \\& \times
\left.
\int_{s_1}^{s_2} ds\, \frac{x_s^{1-{2 \nu}}}{\sqrt{2 x_s-x_s^2-\xi^2}}
\right.
\,+
\\&
\theta(x<\xi)
\frac{1}{\pi}
\int_{0}^1 \frac{dy}{y}
\left[
\left(
1-y \frac{\partial}{\partial y} \right) Q_{2 \nu}(y,t)
\right]
  \\& \times
\left.
\int_{s_1}^{s_3} ds \frac{x_s^{1-2 \nu}}{\sqrt{2 x_s-x_s^2-\xi^2}}
\right.
\,.
\end{split}
\label{Hk_main}
\end{equation}
Here
$x_s= 2 \frac{x- \xi s}{(1-s^2)y}$.
$s_i$, ($i=1,...\,4$)
stand for the four roots of the equation
$2 x_s -x_s^2-\xi^2=0$
given by the following expressions:
\begin{equation*}
\begin{split}
& s_1=\frac{1}{y}
\left(
\mu  - \sqrt{ \left( 1 - x\,y \right) \,\left( 1 + {\mu }^2 \right)-(1 - y^2 )
}
\right); \\&
s_2=\frac{1}{y}
\left(
\mu  + \sqrt{ \left( 1 - x\,y \right) \,\left( 1 + {\mu }^2 \right)-(1 - y^2 )
}
\right);
\\&
s_3=\frac{1}{y}
\left(
\lambda  - \sqrt{ \left( 1 - x\,y \right) \,\left( 1 + {\lambda }^2 \right)-(1 - y^2 )
}
\right);
\\&
s_4=\frac{1}{y}
\left(
\lambda  + \sqrt{ \left( 1 - x\,y \right) \,\left( 1 + {\lambda }^2 \right)-(1 - y^2 )
}
\right),
\end{split}
\end{equation*}
where we have introduced the notations:
\begin{equation*}
\mu= \frac{1-\sqrt{1-\xi^2}}{\xi}; \ \ \ \ \lambda=\frac{1}{\mu}\;.
\end{equation*}
$y=y_0$ and $y=\frac{1}{y_1}$ are two solutions of the equation:
$s_1=s_2$:
\begin{equation*}
y_0=
\frac{x\,\left( 1 + {\mu }^2 \right) }{2} +
{\sqrt{  \frac{x^2\,{\left( 1 + {\mu }^2 \right) }^2}{4}-{\mu }^2 }};
\end{equation*}
\begin{equation*}
y_1=
\frac{x\,\left( 1 + {\lambda }^2 \right) }{2} +
{\sqrt{  \frac{x^2\,{\left( 1 + {\lambda }^2 \right) }^2}{4}-{\lambda }^2 }}.
\end{equation*}

Here we give a short summary of the main features of dual parametrization
approach:
\begin{itemize}
\item Once dual parametrization is used the corresponding
GPD satisfies all general constrains such as the forward limit and the  polynomiality
condition in their full form.

\item At the leading order the scale dependence of forward-like functions
$Q_{2 \nu}(x,t)$
is given by the usual DGLAP evolution equation.

\item The forward-like function
$Q_0(x,t=0) \equiv Q_0(x) $
is expressed through the singlet forward parton distribution
$q(x)+ \bar{q}(x) $
as:
\begin{equation}
Q_0(x)=q(x)+\bar{q}(x)- \frac{x}{2} \int_x^1   \frac{dy}{y^2} \, (q(y)+ \bar{q}(y)).
\label{Q0final}
\end{equation}

\item The basic mechanical characteristics of a hadron
(hadron momentum and angular momentum fractions carried by quarks, radial distribution of
forces experienced by quarks in a hadron) are
contained in the lowest forward-like functions
$Q_0(x)$, $Q_2(x)$.
\end{itemize}

\section{GPD quintessence and Abel tomography }
\label{Sec_quintessence}

The amplitudes of hard exclusive processes are given
by the convolutions of GPDs with the perturbative kernel.
The leading order DVCS amplitude can be expressed through
the corresponding elementary amplitude:
\begin{equation}
A(\xi,t)= \int_0^1 dx H(x,\xi,t) \left[
\frac{1}{\xi-x-i \epsilon} - \frac{1}{\xi+x-i \epsilon}
\right]\,.
\label{DVCSamp}
\end{equation}
In the framework of dual parametrization of GPDs it
turns out possible to specify what part of information
on GPDs can be reconstructed from the known amplitudes of
hard exclusive processes.
In \cite{Tomography} the explicit formulae relating
the special combination of forward-like functions
$Q_{2 \nu}$
to the amplitude of hard exclusive process were derived.

The crucial role in this issue is played by the
so-called
{\em GPD quintessence function}
\cite{Tomography,Polyakov:2007rw,Polyakov:2008xm}:
$$
N(x,t)=\sum_{\nu=0}^\infty x^{2 \nu} Q_{2 \nu}(x,t)\,.
$$
GPD quintessence function
is of particular importance since, according to the result of
\cite{Tomography},
the real and the imaginary parts of the standard amplitude
(\ref{DVCSamp})
are expressed in terms of this function:
\begin{equation}
\text{Im} A(\xi,t)=
\int_{\frac{1- \sqrt{1-\xi^2}}{\xi}}^1
\frac{dx}{x} \, N(x,t)
\left[
\frac{1}{\sqrt{\frac{2x}{\xi}-x^2-1}}
\right]\,; 
\label{ImA_Dual}
\end{equation}
\begin{equation}
\begin{split}
& \text{Re} A(\xi,t)=  2 D(t)+
 \int_0^{\frac{1- \sqrt{1-\xi^2}}{\xi}}
\frac{dx}{x} \, N(x,t) \times \\&
\left[
\frac{1}{\sqrt{1- \frac{2x}{\xi}+x^2}}+
\frac{1}{\sqrt{1+ \frac{2x}{\xi}+x^2}}-
\frac{2}{\sqrt{1 +x^2}}
\right]+ \\&
\int_{\frac{1- \sqrt{1-\xi^2}}{\xi}}^1
\frac{dx}{x} \, N(x,t)
\left[
\frac{1}{\sqrt{1+ \frac{2x}{\xi}+x^2}}
-
\frac{2}{\sqrt{1 +x^2}}
\right]\,.
  \end{split}
  \label{ReA_Dual}
\end{equation}
Here $D(t)$ stands for the so-called
$D$-form factor which actually is the subtraction
constant in the dispersion relation in
$\xi$
plane for the amplitude
$A(\xi,t)$
\cite{Anikin:2007yh,Diehl:2007ru}.
%
%
The following expression for the $D$-form factor was derived in
\cite{Tomography}:
\begin{equation}
\begin{split}
D(t)=&
\int_0^1 \frac{dx}{x}Q_0(x,t)
\left(
\frac{1}{\sqrt{1+x^2}}
-1
\right)+ \\&
\int_0^1 \frac{dx}{x}
\left[
N(x,t)-Q_0(x,t)
\right]
\frac{1}{\sqrt{1+x^2}}\,.
\end{split}
\label{D_form_factor}
\end{equation}

Moreover, the GPD quintessence function
$N(x,t)$
can be recovered from the known imaginary
part of the amplitude
(\ref{DVCSamp}).
The corresponding procedure  was described in
\cite{Tomography,Moiseeva:2008qd}.
With the help of Joukowski conformal map
it turns out possible to present
the relation between the imaginary
part of the amplitude and
$N(x,t)$
(\ref{ImA_Dual})
in the form of the so-called Abel integral equation
\cite{Moiseeva:2008qd}.
This equation can be easily inverted
yielding the following relation for
$N(x,t)$
\cite{Tomography}:
\begin{equation}
\begin{split}
N(x,t)=&
\frac{2}{\pi}\,
\frac{x(1-x^2)}{(1+x)^{\frac{3}{2}}}
\int_{\frac{2x}{1+x^2}}^1
\frac{d \xi}{\xi^{\frac{3}{2}}}
\frac{1}{\sqrt{\xi- \frac{2x}{1+x^2}}}
\times
\\&
\left\{
\frac{1}{2}
\text{Im} A(\xi,t) -
\xi \frac{d}{d \xi}
\text{Im} A(\xi,t)
\right\}\,.
\end{split}
\label{N(x)main}
\end{equation}

Thus the information on GPD which can be recovered
from the amplitude of a hard exclusive process (at fixed hard scale)
is encoded in the GPD quintessence function
$N(x,t)$
and the value of $D$-form factor
$D(t)$.
The formula
(\ref{N(x)main})
allows to restore GPD quintessence function
$N(x,t)$
for
$x \ne 0$
from
the measured imaginary part of the amplitude.
In what follows we are going to show that
$N(x,t)$
may obtain additional
$\delta$-function like
contributions having a support at
$x=0$
which are overlooked by Abel tomography
procedure described above. These contributions affect
the value of $D$-form factor giving additional contribution
to (\ref{D_form_factor}).


\section{Polynomiality and the $D$-term }
\label{Poly_and_D}

The polynomiality condition for Mellin moments of the generalized parton distribution
implies that
\cite{pioneers}:
\begin{equation}
\begin{split}
&\int_{-1}^1 dx \, x^N H(x,\xi,t)= \\&
  h_0^{(N)}(t)+h_2^{(N)}(t) \xi^2+...+h_{N+1}^{(N)}(t) \xi^{N+1}  \ \ \ (N- \text{odd})\, .
\end{split}
\label{Mellin_M}
\end{equation}
In the framework of the dual parametrization
the set of coefficients
$h_k^{(N)}$
can be calculated
from the partial wave decomposition of Mellin moments that follows from
(\ref{FSGPD})
\cite{Polyakov:1998ze}:
\begin{equation}
\begin{split}
  & \int_{-1}^1 dx \, x^N H(x,\xi,t)=
  \xi^{N+1} \sum_{n=1 \atop \text{odd}}^N
  \sum_{l=0 \atop \text{even}}^{n+1}
  B_{nl}(t)
  P_l \left( \frac{1}{\xi} \right) \times \\&
  \frac{\Gamma(\frac{3}{2}) \Gamma(N+1) (n+1)(n+2)}{2^N \Gamma(\frac{N-n}{2}+1)
  \Gamma(\frac{N+n}{2}+\frac{5}{2})}  \ \ \ (N- \text{odd})\, .
\end{split}
  \label{Mellin_Dual}
\end{equation}
Hence, for even
$k \le N+1$:
\begin{equation}
\begin{split}
  & h_k^{(N)}(t)=\sum_{n=1 \atop \text{odd}}^N
  \sum_{l=0 \atop \text{even}}^{n+1}
  B_{nl}(t)
 \frac{1}{(N+1-k)!} P_l^{(N+1-k)} \left( 0 \right) \times \\&
  \frac{\Gamma(\frac{3}{2}) \Gamma(N+1) (n+1)(n+2)}{2^N \Gamma(\frac{N-n}{2}+1)
  \Gamma(\frac{N+n}{2}+\frac{5}{2})} \,.
\end{split}
  \label{hk_dual}
\end{equation}
A special attention is to be payed to the coefficients
$h^{(N)}_{N+1}$
which are generated by the so-called $D$-term
\cite{Polyakov:1999gs}.
An important advantage of dual parametrization of GPDs
comparing {\it e.g.} to double distribution parametrization
is that the $D$-term is its natural ingredient.
The coefficients
$d_n(t)$
of the Gegenbauer expansion of the $D$-term
\begin{equation*}
\begin{split}
& D(z,t)= \\&(1-z^2)
\left[
d_1(t) C_1^{\frac{3}{2}}(z)+ d_3(t) C_3^{\frac{3}{2}}(z)+d_5(t)
C_5^{\frac{3}{2}}(z)+...
\right]\,
\end{split}
\end{equation*}
can be computed with the help of the following generating function
\cite{Tomography}:
\begin{equation*}
\begin{split}
\sum_{n=1 \atop \text{odd}}^{\infty}
& d_n(t) \alpha^n= \\&
\frac{1}{\alpha}
\int_0^1 \frac{dz}{z}
\sum_{\nu=0}^\infty
(\alpha z)^{2 \nu}
Q_{2 \nu}(z,t)
\left(
\frac{1}{\sqrt{1+\alpha^2 z^2}}
- \delta_{\nu 0}
\right).
\end{split}
\end{equation*}


Let us now assume that a certain additional artificial $D$-term
$
\theta(1-\frac{x^2}{\xi^2}) \, \delta{D}\left(\frac{x}{\xi},t\right)
$
is appended to GPD
$H(x,\xi,t)$.
$\delta{D}(z,t)$
is supposed to be an arbitrary odd function with respect to variable $z$
having the support $-1 \le z \le 1$.
Its Gegenbauer expansion reads:
\begin{equation}
\delta{D}(z,t)=(1-z^2) \sum_{n=1 \atop \text{odd}}^\infty \delta{d}_n(t)\,
C_n^{\frac{3}{2}}(z).
\label{D_tilde_term}
\end{equation}
In the Mellin moments the $D$-term contributes only to the highest power of
$\xi$:
\begin{equation}
\begin{split}
&\int_{-1}^1 dx \,x^N  \theta \left(1-\frac{x^2}{\xi^2}\right) \,
\delta{D}\left(
\frac{x}{\xi},t
\right) =  \\&
\xi^{N+1} \sum_{n=1 \atop \text{odd}}^N
\delta{d}_n(t) \frac{\Gamma(\frac{3}{2}) \Gamma(N+1) (n+1)(n+2)}{2^N \Gamma(\frac{N-n}{2}+1)
  \Gamma(\frac{N+n}{2}+\frac{5}{2})} \, .
\end{split}
\label{Dterm_cont_to_mellin}
\end{equation}
The $D$-term contribution to the $N$-th Mellin moment
(\ref{Dterm_cont_to_mellin})
can be incorporated into the general form
(\ref{Mellin_Dual})
employed in the framework of dual parametrization
with the help of the following change of the form factors
$B_{(2 \nu-1) \,  0}= \int_0^1 dx x^{2 \nu-1} Q_{2 \nu}(x)$ ($\nu \ge 1$):
\begin{equation}
\begin{split}
&B_{(2 \nu-1) \; 0}(t) \longrightarrow B_{(2 \nu-1) \; 0}(t)+\delta{d}_{2 \nu-1}(t);
\\&
 Q_{2 \nu}(x,t) \longrightarrow
  Q_{2 \nu}(x,t)-\delta{d}_{2 \nu-1}(t) \frac{2}{(\nu-1)!} \, \delta^{(2 \nu-1)}(x);
\end{split}
\label{Dterm_add_Q}
\end{equation}
Hence we can conclude that:
\begin{itemize}
\item The initial implicit assumption of dual parametrization approach
that the functions
$Q_{2 \nu}(x,t)$
belong to the class of smooth functions seems to
be too restrictive.

\item Instead one has to consider the functions
$Q_{2 \nu}(x,t)$
with
$\nu>0$
as generalized functions. In fact this is quite natural since
the functions
$Q_{2 \nu}$
are not directly measurable.
The physical (observable) quantities are always expressed through
Mellin convolutions of forward-like functions
$Q_{2 \nu}(x,t)$.

\end{itemize}

Let us now turn to the GPD quintessence function
$N(x,t)$.
Once the
$\delta{D}(t)$
term is added to GPD
$N(x,t)$
receives additional singular contribution
with the point-like support:
\begin{equation}
 \delta{N}(x,t)= -\sum_{n=1 \atop \text{odd}}^{\infty}
 x^{n+1} \, \delta{d}_n(t) \frac{2}{n!} \, \delta^{(n)}(x)\, .
\label{Nsing_C}
\end{equation}
Clearly the non-zero contribution to observable
quantities may come only from the convolution
of
$\delta N(x)$
with a kernel having a
$1/x$
singularity at
$x=0$.
Hence
(\ref{Nsing_C})
does not affect the value of
$\text{Im} A(\xi,t)$.
This means that the presence of the contribution of the type
(\ref{Nsing_C})
can not be revealed with the help of the method of
Abel tomography.
As for the real part of the amplitude one can check
that the convolution of
$\delta{N}(x,t)$
with the kernel singular at
$x=0$
appearing in second item of
(\ref{D_form_factor})
results in the explicit contribution
to the $D$-form factor.

Thus, in variance with statement of
\cite{Tomography,Polyakov:2007rw},
we conclude that the correct value of the $D$-form factor
can not be computed just from known
$Q_0(x,t)$
and
$N(x,t)$
recovered with the help of Abel tomography method
(\ref{N(x)main})
since this method overlooks the singular contributions
of type
(\ref{Nsing_C}).
The missing information on the $D$-form factor
can be obtained from the direct measurement
of the real part of the amplitude.

Let us also point that the singular contribution
(\ref{Nsing_C})
does not alter the value of $J$ ($J>0$) Mellin moments of
GPD quintessence function which, according to
the result of
\cite{Tomography},
\cite{Polyakov:2007rw}
are related to the contributions of states with
fixed angular momentum in $t$-channel:
\begin{equation}
\int_0^1 dx \, x^{J-1} N(x,t)=
\frac{1}{2} \int_{-1}^1 dz \, \frac{\Phi_J(z,t)}{1-z}.
\end{equation}
Here
$\Phi_J(z,t)$
stands for the distribution amplitude corresponding to two
quark exchange in the $t$-channel with fixed angular momentum $J$.

\section{On the expansion of GPD around $\xi=0$ to the order $\xi^4$}
\label{Sec_power_exp}

Starting from
(\ref{Hk_main})
one can construct the systematic expansion
of the GPD
$H(x, \xi, t)$
in powers of small
$\xi$
for the fixed value of
$x>\xi$.
In
\cite{Tomography} such an expansion was constructed to the order
$\xi^2$.
Here we present the result up to the order
$\xi^4 \,$.
According to
(\ref{H_dual_through_Qk})
for
$x>\xi$ ($\xi \ge 0$):
\begin{equation}
\begin{split}
& H(x, \xi, t) = \\&
\frac{1}{2} \, H^{(0)}(x,  \xi,  t)+
\frac{1}{2} \, \xi^2 H^{(1)}(x,  \xi,  t)+
\frac{1}{2} \, \xi^4 H^{(2)}(x,  \xi,  t)+  ...
\end{split}
\label{Dual_small_xi_exp}
\end{equation}

The result for
$H^{(0)}(x,t)$
up to the order
$\xi^4$
reads:
\begin{equation}
\begin{split}
& H^{(0)}(x,\xi,t)=
Q_0(x,t)+
\frac{\sqrt{x}}{2}
\int_x^1
\frac{dy}{y^{\frac{3}{2}}}
Q_0(y,t)+ \\&
\xi^2
\left[
\frac{x^2-1}{4x} \frac{\partial }{\partial x}Q_0(x,t)+
\right.
\\&
\left.
\frac{1}{32} \int_x^1 dy Q_0(y,t)
\left\{
\frac{1}{y} \left(
3\,{\sqrt{\frac{x}{y}}} + 3\,{\sqrt{\frac{y}{x}}}
\right)+
\right.
\right.
\\&
\left.
\left.
\frac{1}{y^3} \left(
3\,{\sqrt{\frac{y}{x}}} - {\left( \frac{y}{x} \right) }^{\frac{3}{2}}
\right)
\right\}
\right]+ \\&
\xi^4
\left[
-\frac{(1-x^2)(9x^2+3)}{64 x^3 }
\frac{\partial }{\partial x}Q_0(x,t) \right. + \\&
\frac{2x(1-x^2)^2}{64 x^3}
\frac{\partial^2 }{\partial x^2}Q_0(x,t)+  \\&
\frac{1}{64}
\int_x^1 dy Q_0(y,t)
\left.
\left\{
\frac{1}{y} \left(  \frac{105}{32}{\sqrt{\frac{x}{y}}} + \frac{45}{16}{\sqrt{\frac{y}{x}}} +
  \frac{45}{32}{\left( \frac{y}{x} \right) }^{\frac{3}{2}} \right)+
\right.
\right.
\\&
\left.
\left.
\frac{1}{y^3} \left(
\frac{45}{16}{\sqrt{\frac{y}{x}}} - \frac{3}{8}{\left( \frac{y}{x} \right) }^{\frac{3}{2}} +
  \frac{9}{16} {\left( \frac{y}{x} \right) }^{\frac{5}{2}}
\right)+
\right.
\right.
\\&
\left.
\left.
\frac{1}{y^5} \left(
\frac{45}{32}{\left( \frac{y}{x} \right) }^{\frac{3}{2}} +
  \frac{9}{16}{\left( \frac{y}{x} \right) }^{\frac{5}{2}} -
  \frac{15}{32} {\left( \frac{y}{x} \right) }^{\frac{7}{2}}
\right)
\right\}
\right]+O(\xi^6) \, .
\end{split}
\label{HQ0}
\end{equation}

The result for
$H^{(1)}(x,\xi,t)$
to the order
$\xi^2$
reads:
\begin{equation}
\begin{split}
& H^{(1)}(x,\xi,t)=
 \frac{1}{4} Q_2(x,t)+ \\&
\frac{3}{32}
\int_x^1
dy Q_2(y,t)
\frac{1}{y}
\left(
\frac{1}{2}{\sqrt{\frac{x}{y}}} + {\sqrt{\frac{y}{x}}} +
  \frac{5}{2} {\left( \frac{y}{x} \right) }^{\frac{3}{2}}
\right) + \\&
\xi^2
\left[
\frac{1+x^2}{16 x^2}Q_2(x,t)-
\frac{1-x^2}{16 x} \frac{\partial }{\partial x} Q_2(x,t)+
\right. \\&
\left.
\frac{1}{1024}
\int_x^1
dy
Q_2(y,t) \times
\right. \\&
\left.
\left\{
\frac{1}{y}
\left(
35\,{\sqrt{\frac{x}{y}}} + 45\,{\sqrt{\frac{y}{x}}} +
45\,{\left( \frac{y}{x} \right) }^{\frac{3}{2}} +
  35\,{\left( \frac{y}{x} \right) }^{\frac{5}{2}}
\right)+
\right.
\right. \\&
\left.
\left.
\frac{1}{y^3} \left(
15\,{\sqrt{\frac{y}{x}}} - 3\,{\left( \frac{y}{x} \right) }^{\frac{3}{2}} +
  9\,{\left( \frac{y}{x} \right) }^{\frac{5}{2}} + 75\,{\left( \frac{y}{x} \right) }^{\frac{7}{2}}
\right)
\right\}
\right]+\\&
O(\xi^4)\, .
\end{split}
\label{HQ1}
\end{equation}

Finally the result for
$H^{(2)}(x,\xi,t)$
to the order
$\xi^0$
reads:
\begin{equation}
\begin{split}
& H^{(2)}(x,\xi,t)=
\frac{1}{16}Q_4(x,t)+
\frac{5}{1024}
\int_x^1
dy \,
Q_4(y,t)
\frac{1}{y} \times \\&
\left(
\frac{7}{4}{\sqrt{\frac{x}{y}}} + 3\,{\sqrt{\frac{y}{x}}} +
  \frac{9}{2}{\left( \frac{y}{x} \right) }^{\frac{3}{2}} + 7\,{\left( \frac{y}{x} \right) }^{\frac{5}{2}} +
  \frac{63}{4}{\left( \frac{y}{x} \right) }^{\frac{7}{2}}
\right)+ \\& O(\xi^2) \, .
\end{split}
\label{HQ2}
\end{equation}

Several notes are in order:
\begin{itemize}
\item
One can check that exactly the same expansion is valid for non-singlet
($C=-1$) GPDs. Clearly in this case one has to use the set of non-singlet
forward-like functions.

\item An important property of the expansion of the GPD in powers of
$\xi$
(\ref{Dual_small_xi_exp}), (\ref{HQ0}), (\ref{HQ1}), (\ref{HQ2})
is that up to the particular order
$\xi^{2 \mu}$
it involves only a finite number of functions
$Q_{2 \nu}(x)$
with
$\nu \le \mu$
({\it e.g.} to the order
$\xi^4$
only
$Q_0$, $Q_2$
and
$Q_4$
are relevant). This allows to invert this expansion
and to express the set of functions
$Q_{2 \nu}$
through
GPDs
for various phenomenological parametrizations of GPDs. This problem
is addressed in the next Section.

\item Let us assume the small $x$ behavior for the forward-like
functions to be the following:
$Q_{2 \nu}(x) \sim  \frac{1}{x^{2 \nu+\alpha}}$,
where the power
$\alpha$
governs the small $x$
behavior of forward quark distribution ($\alpha<2$).
Then with the help of explicit calculation one can check that
the $N$-th
Mellin moments of the appropriate coefficients of small
$\xi$
expansion
(\ref{Dual_small_xi_exp})
will produce the values
of
$h_{k}^{(N)}$
for
$k \le N-1$
coinciding with those obtained from
(\ref{hk_dual}).
However, the result for the coefficient at the highest power of
$\xi$
of $N$-th Mellin moment
$h_{N+1}^{(N)}$
obtained from the expansion
(\ref{Dual_small_xi_exp})
does not coincide with the value
(\ref{hk_dual})
if
$\alpha \ge 0$.
\end{itemize}

\section{Expressions for forward-like functions}
\label{Sec_The inv}

Let us assume that the expansion of the singlet (or nonsinglet)
GPD
$H(x, \xi) \equiv H(x, \xi, t=0)$
around
$\xi= 0$
for
$x>\xi$
calculated in the framework of a certain
model is known:
\begin{equation}
H(x, \xi)=
\phi_0(x)+\phi_2(x) \xi^2 + \phi_4(x) \xi^4+O(\xi^6)\,.
\label{Expansion_start}
\end{equation}
Here
$$
\phi_{2 \nu}(x)= \frac{1}{(2\nu)!}\,
\frac{\partial^{2 \nu}  } {\partial \xi^{2\nu}}\,
H(x, \xi)_{\xi=0}\,.
$$
Using this expansion together with
(\ref{Dual_small_xi_exp})
one can try to determine the corresponding functions
$Q_{2 \nu}(x,t=0) \equiv Q_{2 \nu}(x)$
from order to order.

Let us start with the function
$Q_0$.
Clearly, since the GPD calculated in the realistic model has
the correct forward limit
$$
\phi_0(x) \equiv H(x, \xi=0)= \frac{q(x)}{2}\,.
$$
Here
$q(x)$
stands for either for singlet (quark plus antiquark)
or nonsinglet (quark minus antiquark) combination of forward
quark distributions.
Hence for
$Q_0(x)$
we get
\begin{equation*}
q(x)=Q_0(x)+ \frac{\sqrt{x}}{2} \int_x^1 \frac{dy}{y}
\frac{1}{\sqrt{\,y}}\,
Q_0(y)
\end{equation*}
and after inverting it we recover the usual expression
for
$Q_0(x)$:
\begin{equation}
Q_0(x)=q(x)- \frac{x}{2} \int_x^1   \frac{dy}{y^2} \, q(y)\,.
\end{equation}

Let us now consider a more involved case.
In order to express
$Q_2(x)$
one has to invert the following
Mellin convolution:
\begin{equation}
 Q_2(x)+
\frac{3}{8}
\int_x^1
\frac{dy}{y}
\left(
\frac{1}{2} \sqrt{\frac{x}{y}}+
\sqrt{\frac{y}{x}}+
\frac{5}{2}
\left(\frac{y}{x} \right)^{\frac{3}{2}}
\right)Q_2(y)= f_2(x),
\label{Q2eq}
\end{equation}
where
$f_2(x)$
is determined by the
$O(\xi^2)$
term of expansion
(\ref{Expansion_start})
minus the known
$O(\xi^2)$
contribution of
$Q_0(x)$.
Let
$M_{2 \;N}$
and
$f_{2
\;N}$
denote $N$-th
Mellin moments of
$Q_2$
and
$f_2$
respectively:
$$
\int_0^1 dx x^N Q_2(x)=M_{2 \;N}; \ \ \ \ \int_0^1 dx x^N f_2(x)= f_{2
\;N}.
$$
Then the relation between the $N$-th
Mellin moments ($N>0$) of l.h.s. and r.h.s. of
equation (\ref{Q2eq})
reads:
\begin{equation}
M_{2 \;N}
\left(1+
\frac{3}{8}
\frac{8(1+6N+4N^2)}{(-3-2N+12N^2+8N^3)}
\right)=f_{2 \;N}\,.
\label{relation_mellin_mom}
\end{equation}
The relation
(\ref{relation_mellin_mom})
can be easily inverted yielding
the following expression for the
forward-like function
$Q_2(x)$:
\begin{equation*}
Q_2(x)=f_2(x)- \int_x^1 dy f_2(y)
\left(
 \frac{15\,x}{16\,y^2} +
  \frac{3}{8\,y}+\frac{3}{16\,x}
\right).
\end{equation*}
Now using the explicit expression for
$f_2(x)$ we obtain
the following result for
$Q_2(x)$:
\begin{equation}
\begin{split}
& Q_2(x)=
\frac{2(1-x^2)}{x^2} \, q(x)+
\frac{(1-x^2)}{x} \, q'(x)+ \\&
\int_x^1 dy \, q(y) \left(
\frac{-15\,x}{4\,y^4} - \frac{3}{2\,y^3} +
  \frac{5\,x}{4\,y^2}\right)+ \\&
8 \phi_2(x)-
  \int_x^1 dy \,
  \phi_2(y)
\left(
  \frac{15\,x}{2\,y^2} + \frac{3}{y}+\frac{3}{2\,x}
\right).
\end{split}
\label{Q2final}
\end{equation}

In the same manner we can  derive the equation
for
$Q_4(x)$:
\begin{equation}
\begin{split}
& Q_4(x)= f_4(x)- \\&
\int_x^1 dy f_4(y)
\left(
 \frac{35}{256} \frac{y^2}{x^3}+
 \frac{15}{64 } \frac{y}{x^2}+
\frac{45}{128 } \frac{1}{x}+
 \frac{35}{64  } \frac{1}{y}+
 \frac{315}{256} \frac{x}{y^2}
\right)\,,
\end{split}
\end{equation}
where
$f_4(x)$
is determined by the
$O(\xi^4)$
term of expansion
(\ref{Expansion_start})
minus the known
$O(\xi^4)$
contribution of $Q_0(x)$ and $Q_2(x)$.
Using the previously obtained results for
$Q_0(x)$
and
$Q_2(x)$
we derive the following explicit expression for
$Q_4(x)$:
\begin{equation}
\begin{split}
& Q_4(x)=
\left( \frac{49}{8\,x^4} - \frac{17}{8\,x^2} \right) \,\left( 1 - x^2
\right)q(x)+ \\&
\left( \frac{7}{4\,x^3} - \frac{11}{4\,x} \right) \,\left( 1 - x^2 \right)
q'(x)+ \frac{{\left( 1 - x^2 \right) }^2}{2\,x^2} q''(x)-
\\&
\int_x^1dy \,q(y)
\left(
\frac{315\,x}{16\,y^6} + \frac{35}{4\,y^5} -
  \frac{135\,x}{8\,y^4} - \frac{21}{4\,y^3} +
  \frac{27\,x}{16\,y^2}
\right) + \\&
\left(
 \frac{24}{x^2}-40
 \right)
\phi_2(x)+
\frac{8\,\left( 1 - x^2 \right) }{x}\,
\phi_2'(x)-
\\&
\int_x^1 dy
\phi_2(y)
\left(
  \frac{315\,x}{8\,y^4}-\frac{7}{8\,x^3} + \frac{35}{2\,y^3} -
  \frac{135\,x}{4\,y^2} - \frac{21}{2\,y}- \frac{15}{4\,x}
\right)+ \\&
\frac{96}{3} \, \phi_4(x)
- \\&
\int_x^1 dy
\, \phi_4(y)
\left(
\frac{45}{4\,x} + \frac{315\,x}{8\,y^2} + \frac{35}{2\,y} + \frac{15\,y}{2\,x^2} +
  \frac{35\,y^2}{8\,x^3}
\right) \;.
\end{split}
\label{Q4final}
\end{equation}

The expressions for
$Q_{2\nu}$
with
$\nu>2$
can be derived in a completely analogous way although they are very bulky.
Thus we have described the reparametrization procedure
in principle allowing to express any particular forward
like function
$Q_{2 \nu}$
through GPD for a broad class of phenomenological models.

Note, that the starting point for the derivation of
the expressions for
$Q_{2 \nu}(x)$
is the
small
$\xi$
expansion of
$H(x,\xi)$
for
$x>\xi$.
Certainly this expansion is not affected when a
$D$-term $\delta{D} \left(\frac{x}{\xi}\right)$
(\ref{D_tilde_term})
is added to
$H(x,\xi)$.
Hence apart from the smooth part
$Q_{2 \nu}(x)$
with
$\nu>0$
may obtain singular contributions of the type
(\ref{Dterm_add_Q}),
which are certainly overlooked by the reparametrization procedure.




\section{Pion GPDs calculated in the instanton-motivated effective chiral quark model}
\label{Sec_Pion_GPDs}

Now, in order
to illustrate the application of the formalism discussed above,
 we are going to consider a certain specific model.
As an example we have chosen the so-called instanton motivated
effective quark model with
non-local
interactions
\cite{Diakonov:1985eg}
(see Appendix~\ref{app1})
allowing the calculation of pion distribution amplitude (DA),
$2 \pi$ DAs
and pion GPDs at a low
normalization point
\cite{Petrov:1997ve,Petrov:1998kg,Polyakov:1998td,Polyakov:1999gs,Anikin:2000th,Praszalowicz:2001pi,Praszalowicz:2003pr}.

The generalized parton distribution in a pion is defined as Fourier transform
of the matrix element of quark light-cone operator taken
between the pion states:
\begin{equation}
\begin{split}
& \frac{1}{2} \int \frac{d\lambda}{2 \pi} e^{i \lambda x n \cdot \bar{p}}
\langle
\pi^b(p') |
\bar{\psi}^{f'}(-\frac{\lambda n}{2})
\,
 \s{n}
 \,
\psi^f (\frac{\lambda n}{2})| \pi^a(p) \rangle= \\&
\delta^{ab} \delta^{f'f} H^{I=0}(x,\xi,t)+i \epsilon^{a b c} (\tau^c)^{f'f}
H^{I=1}(x,\xi,t).
\end{split}
\end{equation}
Here, as usual,
$\bar{p}=\frac{p+p'}{2}$;
$\Delta=p'-p$;
$t=\Delta^2$.
The skewness parameter
is defined as
$\xi=- \frac{\Delta^+}{2 \bar{p}^+}$.
$H^{I=0}(x, \xi, t)$
and
$H^{I=1}(x, \xi, t)$
stand for isoscalar  and isovector GPDs in a pion.
In the forward limit these GPDs are reduced to the usual
singlet (quark plus antiquark) and
valence (quark minus antiquark) quark
distributions respectively:
\begin{equation}
  H^{I=0}(x, \xi=0, t=0)=\frac{1}{2}\left[
  q(x)+\bar{q}(x)
  \right] ,
\end{equation}
\begin{equation}
 H^{I=1}(x, \xi=0, t=0)=\frac{1}{2}\left[
  q(x)-\bar{q}(x)
   \right].
%
\end{equation}
{\it E.g.} for
$\pi^+$:
\begin{equation}
\begin{split}
& H^{I=0}(x, \xi=0,t=0)= \\&
  \begin{cases}
  & \frac{1}{2} \, [ u^{\pi^+}(x)+ \bar{u}^{\pi^+}(x) ] \ \ \ \ \ \ \ \ \ \; \text{for} \;  x>0 \\
 & \frac{1}{2} \, [ -d^{\pi^+}(-x)- \bar{d}^{\pi^+}(-x) ] \ \ \ \text{for} \;  x<0
 \end{cases} \; .
\end{split}
\end{equation}
and
\begin{equation}
\begin{split}
  &H^{I=1}(x, \xi=0,t=0)= \\&
  \begin{cases}
  & \frac{1}{2}\, [ u^{\pi^+}(x)- \bar{u}^{\pi^+}(x) ] \ \ \ \ \ \ \ \ \ \; \text{for} \;  x>0 \\
 & \frac{1}{2} \, [  -d^{\pi^+}(-x)+\bar{d}^{\pi^+}(-x) ]\ \ \ \text{for} \;  x<0
 \end{cases} \; .
 \end{split}
\end{equation}

In the framework of the effective chiral quark model the
isoscalar pion GPD obtains contributions from three diagrams presented
on
Fig~\ref{Eft_graphs},
while the isovector one only from two first diagrams:
\begin{equation}
\begin{split}
 & H^{I=0}(x, \xi,t)= \mathcal{I}_1(x, \xi, t)+\mathcal{I}_2(x, \xi, t) +
 \mathcal{I}_3(x, \xi, t)\,; \\&
  H^{I=1}(x, \xi,t)= \mathcal{I}_1(x, \xi, t)-\mathcal{I}_2(x, \xi, t) \,.
\end{split}
\label{Hpi}
\end{equation}
One can check that
$\mathcal{I}_1(x, \xi, t)$ is nonzero only if $-\xi \le x \le 1$;
$\mathcal{I}_3(x, \xi, t)$ is nonzero only if $-\xi \le x \le \xi$;
\begin{equation}
\begin{split}
& \mathcal{I}_2(x, \xi, t)=-\mathcal{I}_1(-x, \xi, t); \ \ \
\mathcal{I}_3(-x, \xi,t)= -\mathcal{I}_3(x, \xi,t) \\&
\mathcal{I}_2(x, -\xi, t)=\mathcal{I}_1(x, \xi, t); \ \ \
\mathcal{I}_3(x, -\xi,t)= \mathcal{I}_3(x, \xi,t)
\end{split}
\label{SymProp}
\end{equation}
and hence
$\mathcal{I}_2(-x, \xi,t)$
is nonzero only if
$-1 \le x \le \xi$.

Note that in the nonlinear chiral quark model
\begin{equation}
H^{I=0}(x,\xi,t)=
\begin{cases}
H^{I=1}(x,\xi,t) \ \ \ \text{for} \ \ \ x \ge \xi \\
-H^{I=1}(x,\xi,t) \ \ \ \text{for} \ \ \ x \le -\xi
\end{cases} \, .
\end{equation}
Hence the result
of calculation of forward like functions
$Q_{2 \nu}$
for the case of
isovector quark GPD in a pion will be the
same as that for isoscalar case.
Thus in what follows we are going to consider the case of
isoscalar quark GPD in a pion. For simplicity we set
$t=0$, $m_\pi=0$.
The explicit expression for the relevant contributions
$\mathcal{I}_{1}$, $\mathcal{I}_{2}$, $\mathcal{I}_{3}$
are listed in the
Appendix~\ref{App_I1I2I3_explicit}.

On Fig.~\ref{FigH}
we show the results of calculation of
isoscalar quark GPD in a pion in the framework of nonlocal
chiral quark model for
$t=0$, $m_\pi=0$
and various values
of~$\xi$.



\begin{figure*}
 \begin{center}
  \epsfig{figure= 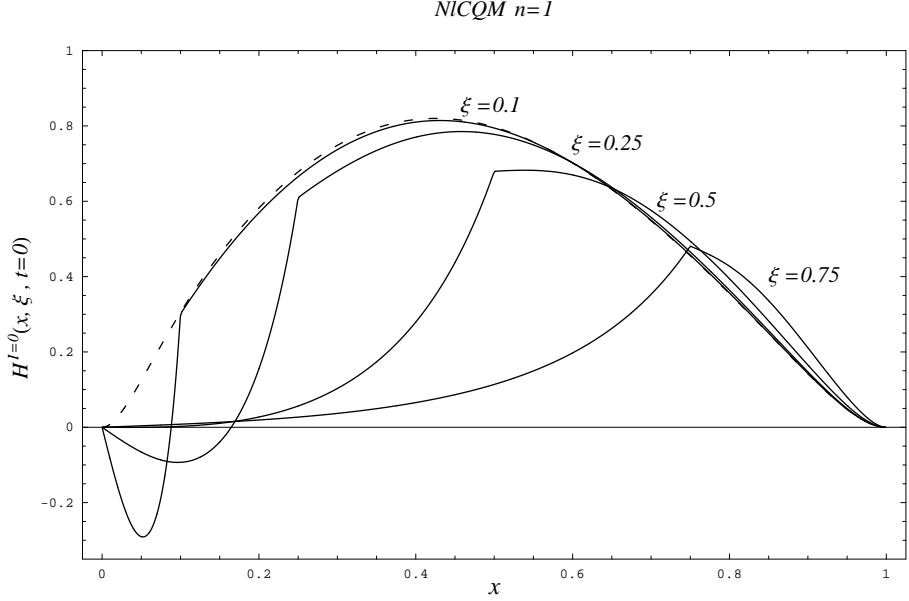 , height=8cm}
\caption{Quark isoscalar GPD
calculated in the nonlocal chiral quark model (NlCQM) for
various values of
$\xi$.
With the dashed line we show $H^{I=0}(x,\xi=0,t=0) \equiv
\frac{1}{2} \left[ q(x)+\bar{q}(x) \right]$.
}
\label{FigH}
\end{center}
\end{figure*}

\section{Calculation of $Q_2(x)$ and $Q_4(x)$ and
checking of the polynomiality condition}
\label{Sec_Q2_Q4_calc}

In this section we illustrate
the application of the reparametrization procedure
described in Sect.~\ref{Sec_The inv}.
Having in hands an analytic result
(\ref{Hpi}),
(\ref{I1_explicit})
for quark isoscalar GPD in a pion
$H(x,\xi)$
for
$x \ge \xi$
calculated in the framework of non-local chiral
quark model we can
explicitly construct its small $\xi$ expansion
for
$x \ge \xi$
(\ref{Expansion_start}).
Next with the help of
(\ref{Q0final}), (\ref{Q2final}), (\ref{Q4final})
we can compute the forward-like functions
$Q_0(x)$, $Q_2(x)$
and
$Q_4(x)$.
It is extremely instructive to compare these results to
the general form of GPD quintessence function
$N(x)$.
This quantity also can be easily computed in the framework of
chiral quark model.

\begin{figure*}
 \begin{center}
  \epsfig{figure= 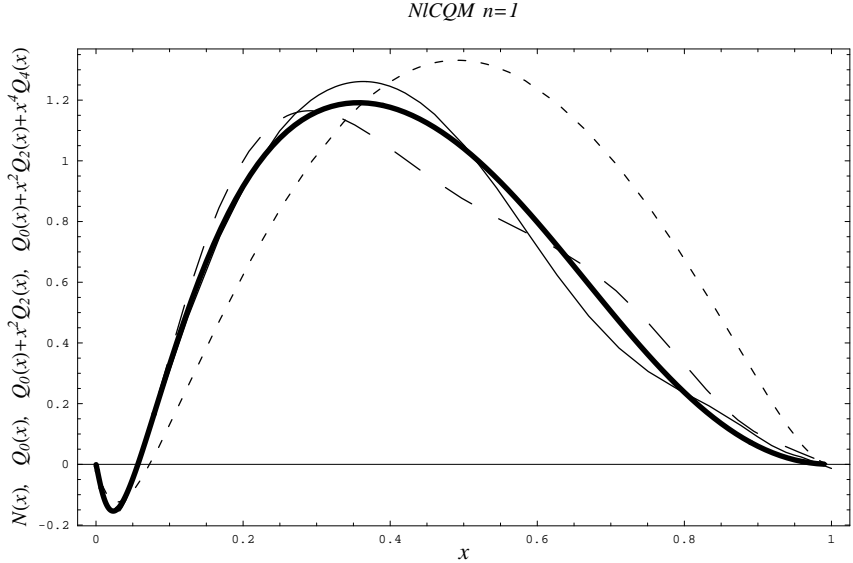 , height=7.5cm}
\caption{We compare our
results for $Q_0(x)$ (short-dashes line),
 $Q_0(x)+x^2Q_2(x)$ (long-dashed line),
 $Q_0(x)+x^2Q_2(x)+x^4Q_4(x)$ (thin solid line) to
$N(x)=\sum_{\nu=0}^\infty x^{2 \nu}Q_{2 \nu}(x)$
(thick solid line).
}
\label{Fig1}
\end{center}
\end{figure*}

\begin{figure*}
 \begin{center}
  \epsfig{figure= 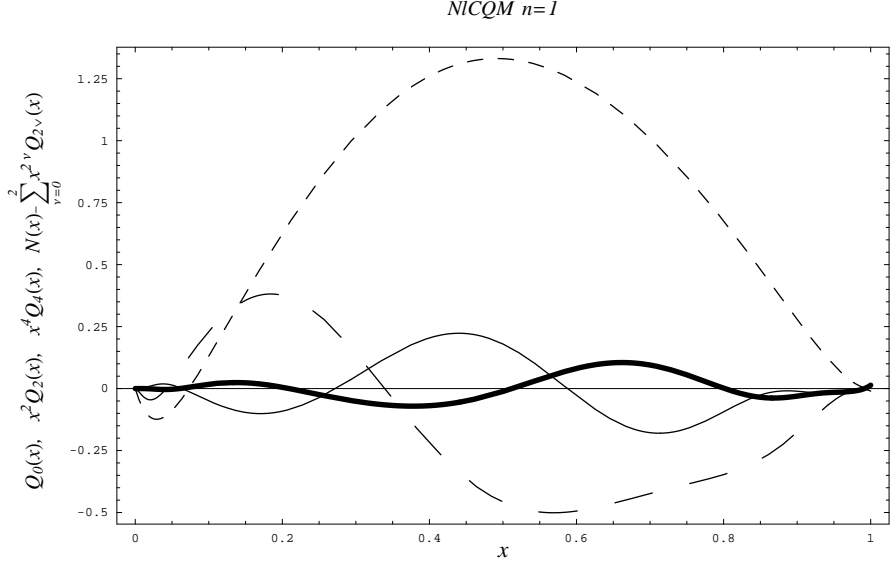 , height=7.5cm}
\caption{We show the results
for $Q_0(x)$ (short-dashed line),
 $x^2Q_2(x)$ (long-dashed line),
 $x^4Q_4(x)$ (thin solid line) and compare them to to
the contribution of
$Q_{2 \nu}(x)$
with
$\nu>2$ into $N(x)$:
$N(x)-\sum_{\nu=0}^2 x^{2 \nu}Q_{2 \nu}(x)=x^6Q_6(x)+...$
(thick solid line).
}
\label{Fig2}
\end{center}
\end{figure*}

In Fig.~\ref{Fig1}
we compare the results for
$Q_0(x)$
(short-dashed line),
$Q_0(x)+x^2Q_2(x)$ (long-dashed line),
$Q_0(x)+x^2Q_2(x)+x^4Q_4(x)$ (thin solid line)
calculated in the framework of nonlocal chiral quark
model with the help of
(\ref{Q0final}), (\ref{Q2final}), (\ref{Q4final})
to GPD quintessence function
$N(x)$
reconstructed from the imaginary part of the DVCS amplitude
(\ref{DVCSamp})
with the help of
(\ref{N(x)main}).
On Fig.~\ref{Fig2}
we show the results for individual contributions of
$Q_{0}(x)$,
$Q_{2}(x)$
and
$Q_{4}(x)$
functions into
$N(x)$.
It is interesting to note that in the case of nonlocal chiral
quark model
the functions
$Q_{2 \nu}(x)$
with higher
$\nu$
provide only small corrections to the values of
$N(x)$.
In other words in the model under consideration the
few first terms really dominate in the
expansion of
$N(x)$
into the sum of
$x^{2 \nu} Q_{2 \nu}(x)$
induced by the inversion of small
$\xi$
expansion of GPD.
In fact the same conclusion remains valid
in the case when Radyushkin
DD-model is used as an input for the dual parametrization
of GPDs (the corresponding analysis will be published elsewhere).

Now we are going to address the problem of polynomiality
of GPD
$H(x,\xi)$.
Let us introduce the following notations for the two
sets of coefficients
$h_{k}^{(N)}$:
\begin{equation}
\begin{split}
& \int_{-1}^1 dx x^N H^{
NlCQM}(x,\xi)=
\sum_{k=0 \atop \text{even}}^{N+1} h_k^{(N) \, NlCQM} \xi^k \,;  \\&
\int_{-1}^1 dx x^N H^{
Dual}(x,\xi)=
\sum_{k=0 \atop \text{even}}^{N+1} h_k^{(N) \, Dual} \xi^k \,.
\end{split}
\end{equation}
Here the coefficients
$h_k^{(N) \, NlCQM}$
can be estimated from the known explicit expression for
GPD in a pion calculated in nonlocal chiral quark model.
The coefficients
$h_k^{(N) \, Dual}$
are given by
(\ref{hk_dual}).
They can be calculated using
(\ref{BnL_def})
together with
the set of functions
$Q_{2 \nu}(x)$
obtained with the help of reparametrization procedure.

Note that the quark isoscalar GPD in a pion
$H(x,\xi)$
calculated in nonlocal quark chiral model satisfies the
so-called soft pion theorem:
$$
H(x,\xi=1)=0.
$$
In particular this means that the following condition is
valid for the coefficients
$h_k^{(N) \, NlCQM}$
of its $N$-th Mellin moment:
\begin{equation}\label{SPT}
h_{N+1}^{(N)\, NlCQM}=-\sum_{k=0 \atop \text{even}}^{N-1}h_k^{(N)\, NlCQM}
\,.
\end{equation}
Using our results for the functions
$Q_0$, $Q_2$, $Q_4$
(\ref{Q0final}), (\ref{Q2final}), (\ref{Q4final})
one can check that for
$k=0,2,4;$
and arbitrary odd
$N > k+1$:
$$
h_{k}^{(N)\, Dual}= 2 \int_0^1 dx x^N \phi_k(x)=h_{k}^{(N)\, NlCQM} \,;
$$
and
$$
h_{0}^{(1)\, Dual}=2 \int_0^1 dx\,  x \phi_0(x)= h_{0}^{(1)\, NlCQM}\, ,
$$
while
\begin{equation*}
\begin{split}
&h_{2}^{(1)\, Dual}=2 \int_0^1 dx\,  x \phi_2(x) \ne h_{2}^{(1)\, NlCQM} \
\\&
h_{4}^{(3)\, Dual}=2 \int_0^1 dx \, x^3 \phi_4(x) \ne h_{4}^{(3)\,
NlCQM} \,.
\end{split}
\end{equation*}

Hence the generalized parton distribution
$H(x,\xi)$
calculated in the framework of dual parametrization with
the help of the functions $Q_0$, $Q_2$, $Q_4$
given by
(\ref{Q0final}) , (\ref{Q2final}), (\ref{Q4final})
does not satisfy the soft pion theorem
(\ref{SPT}).
Fortunately this problem can be cured by
adjusting the values of generalized form factors
$B_{10}$, $B_{30}$.
To do that we need just to add a suitable $D$-term. This results in the
following change of functions $Q_2$, $Q_4$
(see Sect.~\ref{Poly_and_D}):
\begin{equation*}
\begin{split}
& Q_{2}(x) \longrightarrow Q_{2}(x)-\delta {d}_1 \, \frac{2}{1!} \, \delta'(x);
\\&
Q_{4}(x) \longrightarrow Q_{4}(x)-\delta {d}_3 \, \frac{2}{3!} \,
\delta^{(3)}(x) \,.
\end{split}
\end{equation*}
The values of the corresponding coefficients can be easily estimated
numerically:
$$
\delta{d}_1 \approx -0.406\,;
\ \ \ \
\delta{d}_3 \approx -0.018 \,.
$$
The value of $\delta{d}_5$ need to adjust the function $Q_6(x)$ is
$\delta{d}_5 \approx -0.003$.

\section{Calculation of the $D$- form factor}
\label{Calc_D_FF}

\begin{figure*}
 \begin{center}
  \epsfig{figure= 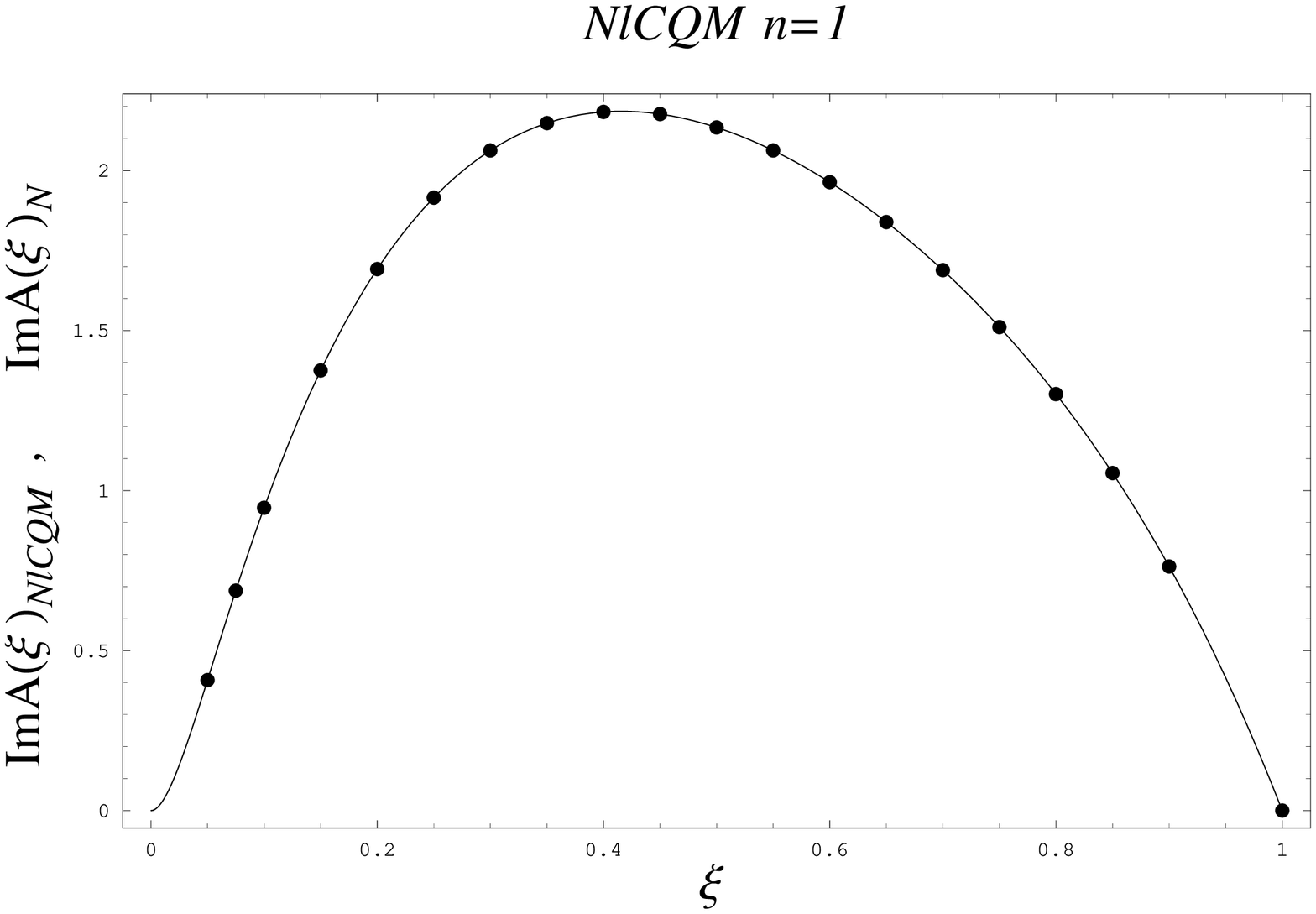 , height=4.91cm}
  \epsfig{figure= 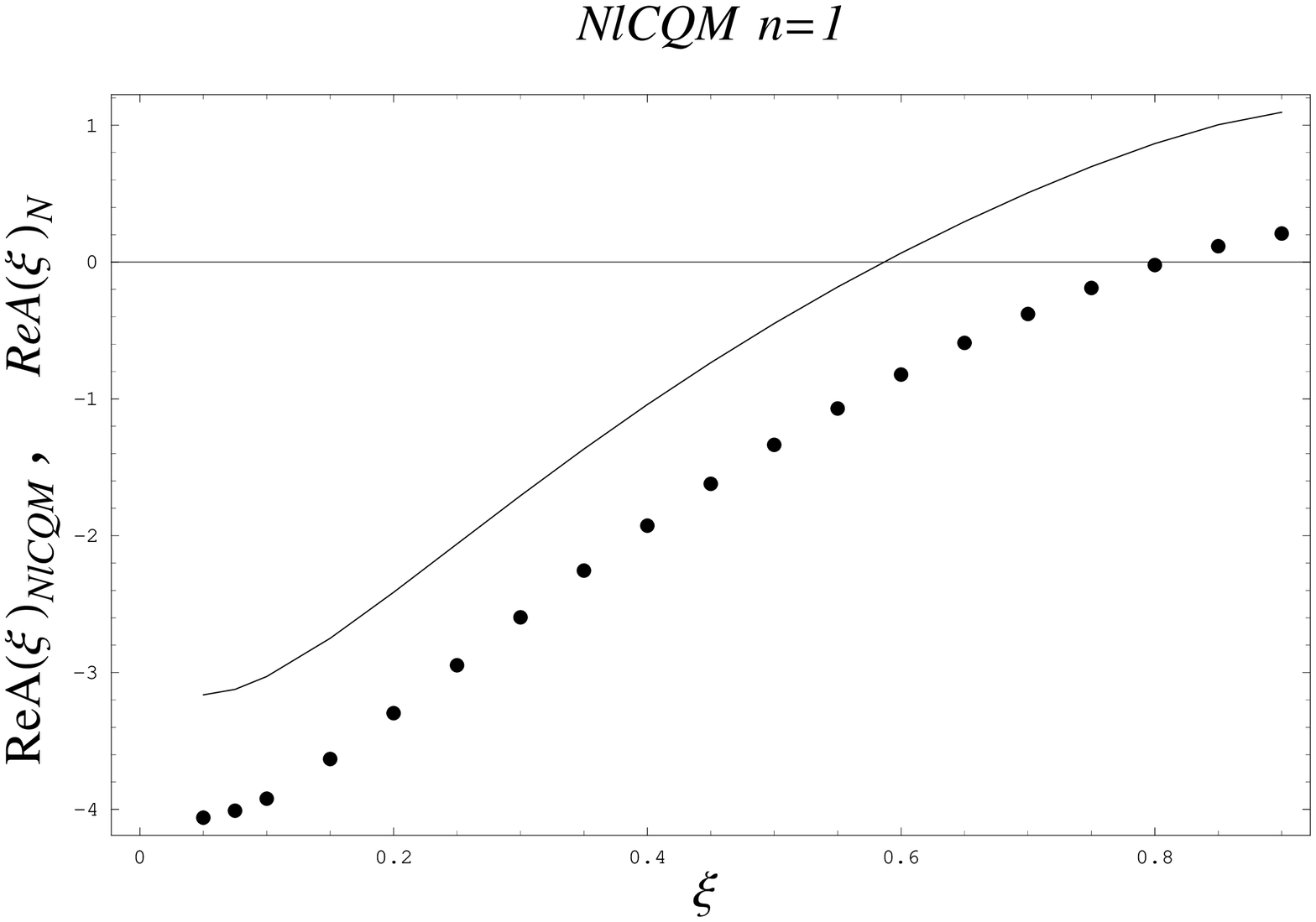 , height=4.91cm}
 \caption{Here we compare the results for the
          imaginary (left panel)  an real (right panel) parts
          of DVCS amplitude calculated in the framework of
          nonlocal chiral quark model (black dots) to
          that obtained from the function
          $N(x)$
          (\ref{ImA_Dual}), (\ref{ReA_Dual})
          (solid lines). The function
          $N(x)$
          is calculated with the help of the inversion formula
          (\ref{N(x)main})
          for which the result for
          $\text{Im}A(\xi)$
          in non-linear chiral quark
          model is used as input.
         }
\label{Fign3}
\end{center}
\end{figure*}

The tomographic procedure of calculation of the
GPD quintessence function
$N(x)$
described in
\cite{Tomography}
is not sensitive to the contributions to
$N(x)$
with the point-like support of the type
(\ref{Nsing_C}),
which do not affect the imaginary part of DVCS amplitude
(\ref{DVCSamp}).
However,
$\delta{N}(x)$
makes an explicit contribution
$$
2 \, \delta{D}= 2 \sum_{k=1 \atop \text{odd}}^\infty  \delta{d}_k
$$
to the $D$-form factor and hence to the real part of DVCS amplitude.
Thus it is extremely edifying to compare the results for
$\text{Re}A(\xi)$
(\ref{ReA_Dual})
calculated with the help of the function
$N(x)$
computed using the inversion formula
(\ref{N(x)main}) with the results of nonlocal chiral quark model
used as an input
to the exact value of
$\text{Re}A(\xi)$
in nonlocal chiral quark model
obtained by the direct calculation of principal value integral in
(\ref{DVCSamp}).
The result is presented on Fig.~\ref{Fign3}.

The imaginary part of DVCS amplitude is accurately reproduced
with the help of the GPD quintessence function computed using
(\ref{N(x)main}).
However, the corresponding real part indeed differs by a certain constant
$2 \, \delta D$ ($\delta D \approx -0.443 $)
from the exact value of
$\text{Re}A(\xi)$
calculated from
(\ref{DVCSamp}).
This difference should certainly be taken as an effect of the singular
term
$\delta{N}(x)$
overlooked by the inversion procedure
(\ref{N(x)main}).
Note that first three terms
$\delta{d}_1$, $\delta{d}_3$, $\delta{d}_5$
(that specify the corrections
(\ref{Dterm_add_Q})
for
$Q_2$, $Q_4$
and
$Q_6$
forward-like functions) calculated in the previous Section
give already a reasonable approximation to
$\delta{D}$
value:
$\delta{d}_1+\delta{d}_3+\delta{d}_5 \approx -0.427$.

Finally,
using the results
$N(x)$,
$Q_0(x)$
and
$\tilde{D}$
we can estimate the complete
$D$-form factor
in nonlocal chiral quark model:
$$
D^{NlCQM}(t=0)= \sum_{k=1 \atop \text{odd}}^\infty
(d_k+\delta{d}_k) \approx -0.523 \,.
$$

It is also extremely instructive to estimate the first
coefficients of the Gegenbauer expansion of the $D$-term
which has clear physical interpretation.
According to the results of
\cite{Polyakov:2002wz,Polyakov:2002yz}
$d_1$
is associated with the forces experienced by the quarks
inside a hadron:
\begin{equation*}
d_1(t=0)= - \frac{m_\pi}{2} \int d^3 r \,
T_{j k}^Q (\vec{r})(r^j r^k- \frac{1}{3} \delta^{jk}r^2),
\end{equation*}
where
$T_{jk}^Q$
is the pion matrix element of the quark stress tensor.
Using our final expression for $Q_2(x)$ we  obtain:
$
d_1(t=0)=  B_{1\,0}+ \delta{d_1}-\frac{1}{2}B_{1\,2} \approx -0.580\,.
$


\section{Conclusions}
\label{concl}

In the framework of dual parametrization of GPDs
we have derived the inversion formulas allowing
to express the forward-like functions
$Q_2$, $Q_4$
through GPDs once GPD is known as a function of
$x$
and
$\xi$.
This reparametrization procedure can be applied
for a broad class of phenomenological models for GPDs.
To provide an example of application of this techniques
we have considered isoscalar GPD in a pion calculated
in the framework of the effective chiral model.
We show that in this model GPD quintessence function
$N(x)$
is with high accuracy saturated by the contributions of the
first few forward-like functions
$Q_{2 \nu}$.
We also argue that the $D$-form factor can not be computed
with the help of the forward-like function
$Q_0(x,t)$
and regular part of GPD quintessence function
$N(x,t)$,
which can be recovered from hard exclusive process amplitude
with the help of Abel tomography method. The reason for this
is that Abel tomography method overlooks the contributions to
$N(x,t)$
having the form of singular generalized function with the
support at $x=0$.
We illustrate this statement in the framework of the
effective chiral model.

\section*{Acknowledgements}

I am grateful to Maxim Polyakov for suggesting this work and
numerous helpful comments and discussions. The work is supported by
the Sofja Kovalevskaja Programme of the Alexander von Humboldt
Foundation and by the Deutsche Forschungsgemeinschaft.

\setcounter{section}{0}
\setcounter{equation}{0}
\renewcommand{\thesection}{\Alph{section}}
\renewcommand{\theequation}{\thesection\arabic{equation}}

\section{Basic facts about effective chiral quark model}
\label{app1}

Pions, which are the Goldstone bosons of spontaneously broken
chiral symmetry, allow the calculations of their properties
with little dynamical input relying upon the chiral structure and
chiral symmetry breaking.
The effective chiral quark model
\cite{Diakonov:1985eg,Petrov:1997ve}
with nonlocal
interaction was
successfully applied to the calculation of
leading twist pion DA,
$2\pi$DAs
and pion GPDs at low normalization point
\cite{Petrov:1997ve,Petrov:1998kg,Polyakov:1998td,Polyakov:1999gs,Anikin:2000th,Anikin:2000rq,Praszalowicz:2001pi,Praszalowicz:2003pr}.

The corresponding effective action (in the momentum space)
reads:
\begin{equation}
\begin{split}
  &S_{\text{eff}}=
\int \frac{d^4 k}{(2 \pi^4)}
\bar{\psi}(k) \, \s{k} \,  \psi(k)- \\&
\int \frac{d^4 p}{(2 \pi^4)} \,
\frac{d^4 k}{(2 \pi^4)}
\bar{\psi}(p)
\sqrt{M_p} \, U^{\gamma_5}(p-k) \,  \sqrt{M_k} \psi(k)\,.
\end{split}
\label{Eff_Th_Action}
\end{equation}
$M_k$
stands for the momentum dependent quark mass
(\ref{Mom_dep_M})
and the matrix
$U^{\gamma_5}$
describes the interaction between quarks and pions:
\begin{equation*}
\begin{split}
&U^{\gamma_5}(x)=e^{\frac{i}{F_\pi} \gamma^5 \tau^a \pi^a(x)}= \\&
1+ \frac{i}{F_{\pi}} \gamma_5 \pi^a(x) \tau^a- \frac{1}{2 F_\pi^2}
\pi^a(x) \pi^a(x)+ ... \,.
\end{split}
\end{equation*}
Hence the effective theory
(\ref{Eff_Th_Action})
contains two type of vertices relevant for the calculation
of matrix elements of twist-two quark operator between pion states:
a Yukawa-type quark-pion vertex and a two-pion quark vertex.

The important ingredient of the model is the momentum dependence
of the quark mass.
In \cite{Praszalowicz:2003pr} this dependence was taken
in the instanton motivated form:
\begin{equation}
M_k=M \left(
\frac{-\Lambda^2}{k^2-\Lambda^2+i \epsilon}
\right)^{2n}
\label{Mom_dep_M}
\end{equation}
(see also
\cite{Praszalowicz:2001pi}
for the corresponding discussion).
Quantities
$\Lambda$
and
$n$
are model parameters.
The parameter
$\Lambda$
in this model is fixed by adjusting the value of pion decay constant
$F_\pi(\Lambda)$
to its physical value.
The dependence of the results on the value of $n$
was reported to be rather weak
\cite{Praszalowicz:2001pi,Praszalowicz:2003pr}.
Following the choice of
\cite{Praszalowicz:2003pr}.
we fix
$n=1$.
For the constituent quark mass at zero momentum
$M=350$~MeV,
$n=1$
and
$F_\pi=93$~MeV
the parameter
$\Lambda$
is to be set to
$\Lambda=1157$~MeV.

\setcounter{equation}{0}

\section{Pion GPDs in nonlocal chiral quark model}
\label{App_I1I2I3_explicit}

\begin{figure*}
 \begin{center}
    \epsfig{figure= 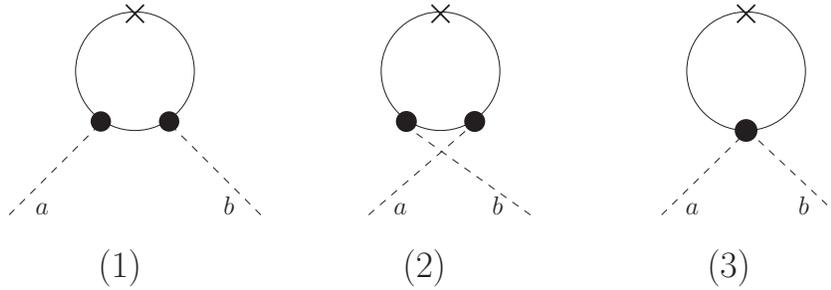 , height=3.8cm}
 \caption{Diagrams of the effective chiral quark theory
(\ref{Eff_Th_Action})
 contributing
 to the matrix element
$\langle
\pi^b(p') |
\bar{\psi}^{f'}(-\frac{\lambda n}{2})
\,
 \s{n}
 \,
\psi^f (\frac{\lambda n}{2})| \pi^a(p)
\rangle$.}
\label{Eft_graphs}
\end{center}
\end{figure*}

Below we list the explicit results of
\cite{Praszalowicz:2003pr}
for the contributions of diagrams
presented on
Fig.~\ref{Eft_graphs}
into pion GPDs
(\ref{Hpi}).
According to the symmetry properties
(\ref{SymProp})
we need expressions only for
$\mathcal{I}_1$
and
$\mathcal{I}_3$
contributions.
For simplicity we set
$t=0$, $m_\pi=0$.

Then the result of
\cite{Praszalowicz:2003pr}
for the integrals
relevant for
$-\xi \le x \le \xi$
reads:
\begin{equation}
\begin{split}
& \mathcal{I}_1(x,\xi,t=0)= \frac{ (-1)^{n+1} N_C M^2}{2 (2 \pi)^2 F_\pi^2} \times \\&
\int_0^\infty d (\kappa_\bot^2)
\sum_{i=1}^{4n+1} \phi_i
\frac{(x-1)^{6n+1} (f_i \, g_i)^n}{ { \prod_{k=1}^{4n+1} }B_{ik} D_{ik}}
h_1^{(a)}(x, \xi),
\end{split}
\end{equation}
where
$N_C=3$
is the number of colors,
$\kappa$
is the scaled variable
($\kappa^\nu=
\frac{k^\nu}{\Lambda}$),
$\mu= \frac{M}{\Lambda}$.
For the definition of factors
$f_i$, $g_i$, $B_{ik}$,  $D_{ik}$
and
$\phi_i$
see
(\ref{factors_we_need}),
(\ref{phi_i}).
\begin{equation}
\begin{split}
  & h_1^{(a)}(x, \xi)= \\&
  \mu^2
\left[
(x-\xi) \frac{b_i^{2n}}{(x-\xi)^{2n}}+
(x+\xi) z_i^{2n}-
(x-1) \frac{d_i^{2n}}{(x+\xi)^{2n}}
\right]+ \\&
\frac{b_i^{2n} d_i^{2n}}{(x+\xi)^{4n}} z_i^{2n}
\left[
(\xi+1) \kappa_\bot^2+ (\xi-x)(1+z_i)
\right].
\end{split}
\end{equation}

For
$-\xi \le x \le \xi$
$\mathcal{I}_3(x,\xi,t=0)$
reads:
\begin{equation}
\begin{split}
&\mathcal{I}_3(x,\xi,t=0)= \\&
\frac{(-1)^{n+1}N_C M^2}{2 (2\pi)^2 F_\pi^2}
\int_0^\infty d(\kappa_\bot^2)
\sum_{i=1}^{4n+1} \phi_i
\left(
(x+\xi)z_i b_i
\right)^n
\times \\&
\frac{(x+\xi)^{2n+1}z_i^{2n}+(x-\xi) b_i^n}{\prod_{k=1}^{4n+1}
B_{ik}}\,.
\end{split}
\end{equation}

For
$\xi \le x \le 1$
only the contribution of
$\mathcal{I}_1$
is non-zero.
It reads:
\begin{equation}
\begin{split}
& \mathcal{I}_1(x,\xi,t=0)= \frac{ (-1)^{n}N_C M^2}{2 (2 \pi)^2 F_\pi^2} \times \\&
\int_0^\infty d (\kappa_\bot^2)
\sum_{i=1}^{4n+1} \phi_i
\frac{(x-1)^{6n+1} (f_i \, g_i)^n}{ { \prod_{k=1}^{4n+1} }F_{ik} G_{ik}}
h_1^{(b)}(x, \xi),
\end{split}
\label{I1_explicit}
\end{equation}
where
\begin{equation}
\begin{split}
  & h_1^{(b)}(x, \xi)= \\&
  \mu^2
\left[
(x-\xi) \frac{g_i^{2n}}{(x-1)^{2n}}+
(x+\xi) \frac{f_i^{2n}}{(x-1)^{2n}}-
(x-1)z_i^{2n}
\right]+ \\&
\frac{f_i^{2n} g_i^{2n}}{(x-1)^{4n}}z_i^{2n}
\left[
\frac{\xi^2-1}{x-1} \kappa_\bot^2+ \frac{\xi^2-x^2}{x-1}(1+z_i)
\right].
\end{split}
\end{equation}

The symbols
$b_i$, $d_i$,
$B_{ik}$, $D_{ik}$
$f_i$, $g_i$, $F_{ik}$, $G_{ik}$
($i,k=1,...,4n+1$)
are defined as:
\begin{equation}
\begin{split}
&
b_i=2 \xi(\kappa_\bot^2+1)+(\xi-X)z_i\,; \\&
d_i=-(1+\xi)(\kappa_\bot^2+1)+(x-1)z_i\,;\\&
B_{ik}=2\xi(\kappa_\bot^2+1)+(\xi-x)z_i+(\xi+x)z_k\,;\\&
D_{ik}=-(1+\xi)(\kappa_\bot^2+1)+(x-1)z_i-(\xi+x)z_k\,;\\&
f_i=-(1+\xi)(\kappa_\bot^2+1)-(\xi+x)z_i \, ; \\&
g_i=(1-\xi)(\kappa_\bot^2+1)-(\xi-x)z_i\, ; \\&
F_{ik}=-(1+\xi)(\kappa_\bot^2+1)-(\xi+x)z_i+(x-1)z_k\, ; \\&
G_{ik}=(1-\xi)(\kappa_\bot^2+1)-(\xi-x)z_i-(x-1)z_k \, .
\end{split}
\label{factors_we_need}
\end{equation}

$z_i$, ($i=1,...,\,4n+1$)
stand for the solutions of the master equation which
controls the position of poles of the fermion propagator:
$$
z^{4n+1}+z^{4n}-\frac{M^2}{\Lambda^2}=0 \;.
$$
The factors
$\phi_i$
are introduced according to:
\begin{equation}
\phi_i= \prod_{k=1 \atop k \ne i}^{4n+1} \frac{1}{z_i-z_k};
\ \ \ \ \
\sum_{i=1}^{4n+1} z_i^m \phi_i =
\begin{cases}
0  & \text{for} \;  0 \le m<4n \\ 1
 & \text{for} \; m=4n
 \end{cases} \; .
\label{phi_i}
\end{equation}


\end{document}